\documentclass[11pt]{article}
\usepackage{amsmath,amssymb,color,graphics,epsfig}
\usepackage{graphicx}
\usepackage{subfigure}

\usepackage{amsfonts}

\newcommand{\be}{\begin{equation}}
\newcommand{\ee}{\end{equation}}
\newcommand{\bea}{\setlength\arraycolsep{2pt} \begin{eqnarray}}
\newcommand{\eea}{\end{eqnarray}}
\newcommand{\nn}{\nonumber}
\newcommand{\mc}{\mathcal}
\newcommand{\mm}{\mathrm}

\def\ft#1#2{{\textstyle{\frac{\scriptstyle #1}{\scriptstyle #2} } }}
\def\fft#1#2{{\frac{#1}{#2}}}

\def\0{{\sst{(0)}}}
\def\1{{\sst{(1)}}}
\def\2{{\sst{(2)}}}
\def\3{{\sst{(3)}}}
\def\4{{\sst{(4)}}}
\def\5{{\sst{(5)}}}
\def\6{{\sst{(6)}}}
\def\7{{\sst{(7)}}}
\def\8{{\sst{(8)}}}
\def\sst#1{{\scriptscriptstyle #1}}

\begin{document}

\begin{flushright}
\end{flushright}

\vspace{25pt}
\begin{center}
{\large {\bf Universal relation for operator complexity}}

\vspace{10pt}
 Zhong-Ying Fan$^{1\dagger}$\\

\vspace{10pt}
$^{1\dagger}${ Department of Astrophysics, School of Physics and Material Science, \\
 Guangzhou University, Guangzhou 510006, China }\\


\vspace{40pt}

\underline{ABSTRACT}
\end{center}
We study Krylov complexity $C_K$ and operator entropy $S_K$ in operator growth. We find that for a variety of systems, including chaotic ones and integrable theories, the two quantities always enjoy a logarithmic relation $S_K\sim \mm{ln}{C_K}$ at long times, where dissipative behavior emerges in unitary evolution. Otherwise, the relation does not hold any longer. Universality of the relation is deeply connected to irreversibility of operator growth.

\vfill {\footnotesize  Email: $^\dagger$fanzhy@gzhu.edu.cn\,,}

\thispagestyle{empty}

\pagebreak

\tableofcontents
\addtocontents{toc}{\protect\setcounter{tocdepth}{2}}




\section{Introduction}

 Time evolution in quantum mechanical systems is generally local and unitary. However, it is also known that many quantum systems have effectively irreversible hydrodynamic descriptions, for example the transport properties. It is a central goal to understand emergence of this thermal behavior in theoretical research on quantum many body systems. Operator growth and the relation to thermalization is an exciting topic is in this direction, for example see \cite{vonKeyserlingk:2017dyr,Nahum:2017yvy,Khemani:2017nda,Rakovszky:2017qit,Gopalakrishnan:2018wyg}.

 Consider an initial operator $\mc{O}_0$ and suppose it can be written as a sum of a few basis vectors in any local basis. Heisenberg evolution of the space of operators, $\mc{O}(t)=e^{iHt}\mc{O}_0 e^{-iHt}$, is described by a set of nested commutators of $\mc{O}_0$ with the Hamiltonian $H$. Evaluation of all these commutators is equivalent to solving dynamics of the operator completely. However, for complex systems, the number of these commutators (or the nonzero coefficients of $\mc{O}(t)$ in any local basis) increases monotonically in the evolution and can blow up exponentially: the initially simple operator $\mc{O}_0$ grows irreversibly into a complex one. Because of exponential size of the problem, a statistical description should emerge for the process. This implies that operator growth should have some form of universality, similar to statistical mechanics. It is of great interests to search universal features of operator dynamics.

 In this paper, we would like to study information quantities, the Krylov complexity ( or K-complexity) \cite{Parker:2018yvk} and the operator entropy (or K-entropy) \cite{Barbon:2019wsy} in operator growth. Previously, the two quantities were examined separately in literature \cite{Caputa:2021sib,Jian:2020qpp,Rabinovici:2020ryf,Dymarsky:2021bjq,Kim:2021okd,Patramanis:2021lkx}. Here we are interested in examining their functional relation in the time evolution. We will show that the two enjoys a logarithmic relation at long times\footnote{In this paper, the base of the logarithm is the constant $e$.}
 \be\label{skchaotic} S_K(t)=\tilde{\eta}\,\mm{ln}{C_K(t)}+\cdots \,,\ee
where dissipative behavior emerges. Here $\tilde{\eta}$ is a positive constant, depending on the choice of dynamics.
We propose that it is up bounded as $\tilde\eta\leq 1$. The relation is universal to all irreversible growth of operators, not necessarily exponential growth, as far as we can check.

The paper is organized as follows. In section 2, we briefly review the recursion method and a quantitative measure for irreversibility of operator growth. In section 3, we study the functional relation between $S_K$ and $C_K$ at initial times. We show that they enjoy a product logarithmic relation $S_K\sim -C_K\log{C_K}$ to leading order. In section 4, we study the long time behavior of $S_K$ and $C_K$ for a variety of systems with emergence of dissipative behaviors\footnote{In this paper, ``dissipative behavior" was referred to the decaying behavior of operator wave functions at long times.}, including chaotic ones and integrable theories. We find that they always enjoy a logarithmic relation $S_K\sim \log{C_K}$ at leading order, irrespective of dynamical details. In section 5, as a comparison, we examine simple systems, which have no emergence of dissipative behaivors and show that in this case, the logarithmic relation does not hold any longer. We conclude in section 6.


\section{Preliminaries: recursion method and irreversibility of operator dynamics }\label{sec2}

For a lattice system with Hamiltonian $H$, the initial operator $\mc{O}_0$ evolves in time according to the Heisenberg equation $\partial_t\mc{O}(t)=i[H\,,\mc{O}(t)]$ or explicitly
\be \mc{O}(t)=e^{iHt}\mc{O}_0 e^{-iHt}=\sum_{n=0}\fft{(it)^n}{n!}\tilde{\mc{O}}_n\,, \ee
where $\tilde{\mc{O}}_n$ stands for nested commutators: $\tilde{\mc{O}}_1=[H,\mc{O}_0]\,,\tilde{\mc{O}}_2=[H,\tilde{\mc{O}}_1]\,,\cdots\,,\tilde{\mc{O}}_n=[H,\tilde{\mc{O}}_{n-1}]\,.$
However, evaluation of these commutators is of great difficult for complex systems. Sometimes it is helpful to thick of the problem as solving a Schr\"{o}dinger equation by taking the operator as a wave function. Define the {\it Liouvillian} $\mc{L}\equiv [H\,,\cdot]$, the operator wave function evolves as
\be\label{obasis} |\mc{O}(t)\rangle=e^{i\mc{L}t}|\mc{O}_0\rangle= \sum_{n=0}\fft{(it)^n}{n!}|\tilde{\mc{O}}_n\rangle\,, \ee
where $|\tilde{\mc{O}}_n\rangle=\mc{L}^n|\mc{O}_0\rangle.$ In this language, one generally needs to introduce a proper inner product in the operator Hilbert space. For example, in the infinite temperature limit, one usually has $\langle A|B\rangle=\mm{Tr}\big(A^{\dag} B \big) $. We refer the readers to \cite{vsmuller} for more details. The physical information about operator growth is essentially encoded in the so-called auto-correlation function, which is defined as
\be C(t):= \langle \mc{O}_0|\mc{O}(t)\rangle=\langle \mc{O}_0| e^{i\mc{L}t}  |\mc{O}_0\rangle\,. \ee
The same information can also be extracted from the {\it moments} $\mu_{2n}$
\be
\mu_{2n}:= \langle \mc{O}_0| \mc{L}^{2n}  |\mc{O}_0\rangle=\fft{d^{2n}}{dt^{2n}}C(-it)\Big|_{t=0}\,.
\ee
Note $C(t)=\sum_{n=0}\fft{\mu_{2n}(it)^{2n}}{(2n)!}=1-\fft{1}{2}\mu_2t^2+\cdots$, implying that the initial growth of operators is determined by the lowest moment $\mu_2$. We will come to this point in the next section.

Another two quantities encoding the same information about the dynamics of operators are the relaxation function $\phi_0(z)$ and the spectral density $\Phi(\omega)$. They are defined as the Laplace transform and the Fourier transform of $C(t)$, respectively
\bea
&& \phi_0(z):= \int_{0}^{+\infty}dt\, e^{-z t} C(t)\,,\nn\\
&& \Phi(\omega):= \int_{-\infty}^{+\infty}dt\, e^{-i\omega t} C(t) \,.
\eea
Note that they (and the moments ) are linearly related to the auto-correlation function. In particular, one has \cite{vsmuller}
 \be \Phi(\omega)=2\lim_{\varepsilon\rightarrow 0}\mm{Re}\big[\phi_0(\varepsilon-i\omega) \big] \,.\ee
 In the remaining of this paper, we will frequenctly switch between these quantities and use the one that is most convenient to discuss the physics we are interested in. The readers should not be confused.

\subsection{The recursion method }
In general, the original operator basis $\{|\tilde{\mc{O}}_n\rangle\}$ is not orthogonal. Just like in ordinary quantum mechanics, one can study dynamics of the operator wave function using different basises, of which a particularly interesting one is orthonormal. This leads to a different approach to operator growth. Let us briefly review it in the following.

Using Gram-Schmidt scheme, one starts with a normalized vector $|\mc{O}_0\rangle$. Then the first vector is given by $|\mc{O}_1\rangle:=b_1^{-1}\mc{L}|\mc{O}_0\rangle$, where $b_1:=\langle \mc{O}_0\mc{L}|\mc{L}\mc{O}_0 \rangle^{1/2}$. For the $n-$th vector, one inductively defines
\bea\label{kbasis}
&&|A_n\rangle:=\mc{L}|\mc{O}_{n-1}\rangle-b_{n-1}|\mc{O}_{n-2}\rangle\,,\nn\\
&&|\mc{O}_n\rangle:=b_n^{-1}|A_n\rangle\,,\quad b_n:=\langle A_n|A_n \rangle^{1/2}\,.
\eea
The output of the above procedure is a set of orthomornal vectors $\{|\mc{O}_n\rangle\}$, called {\it Krylov basis} and a sequence of positive numbers $\{b_n\}$, referred to as {\it Lanczos coefficients}, which have units of energy (in this paper, time is measured in units of the inverse of Lanczos coefficients, for example $1/b_1$).
The physical information encoded in the auto-correlation function or moments can be equivalently extracted from the Lanczos coefficients. However, they are related via nonlinear transformations. In Krylov basis, the Liouvillian is tridiagonal
\be L_{nm}:=\langle O_n|\mc{L}|O_m \rangle=\left(
\begin{array}{cccccccc}
 0 & b_1 & 0  & 0 & \ldots\\
 b_1  & 0 & b_2 & 0 & \ldots\\
 0 & b_2 & 0 & b_3  & \ldots\\
 0 & 0 & b_3 & 0 & \ddots \\
 \vdots & \vdots & \vdots & \ddots & \ddots\\
\end{array}
\right)\,,\ee
whereas the moments are given by
\be \mu_{2n}= \langle \mc{O}_0| \mc{L}^{2n}  |\mc{O}_0\rangle=(L^{2n})_{00} \,.\ee
Some low lying examples are
\be \mu_2=b_1^2\,,\quad \mu_4=b_1^4+b_1^2b_2^2\,,\quad \cdots \,.\ee
General transformations between these two sets of constants can be found in \cite{vsmuller} (or Appendix A of \cite{Parker:2018yvk}). The nonlinear information coding of Lanczos coefficients plays an indispensable role in demonstrating universality of operator growth \cite{Parker:2018yvk}.

In Krylov basis, evolution of the operator $\mc{O}(t)$ can be formally written as
\be\label{newbasis} |\mc{O}(t)\rangle:=\sum_{n=0}i^n\varphi_n(t)|\mc{O}_n\rangle\,,  \ee
where $\varphi_n(t):=i^{-n}\langle \mc{O}_n|\mc{O}(t)\rangle$ is a discrete set of wave functions. The Heisenberg evolution of $\mc{O}(t)$ gives rise to a discrete set of equations
\be\label{varphi} \partial_t\varphi_n=b_n\varphi_{n-1}-b_{n+1}\varphi_{n+1} \,,\ee
subject to the boundary condition $\varphi_n(0)=\delta_{n0}$ and $b_0=0=\varphi_{-1}(t)$ by convention. This defines a one-dimensional quantum mechanical problem on a half chain. This uniquely determines the wave functions $\varphi_n(t)$ for a given set of Lanczos coefficients. Since the auto-correlation function is simply $C(t)=\varphi_0(t)$, it is completely equivalent to the Lanczos coefficients.

Taking the Laplace transform of (\ref{varphi}), one finds
\be z\phi_n(z)-\delta_{n0}=b_n \phi_{n-1}(z)-b_{n+1}\phi_{n+1}(z) \,,\ee
where $\phi_n(z)$ is the Laplace transform of $\varphi_n(t)$. The relation gives the continued fraction representation of the relaxation function
\be
\phi_0(z)=\dfrac{1}{z+\dfrac{b_1^2}{z+\dfrac{b_2^2}{  z+\ddots  } } } \,.
\ee
For finite chains, the recursion method naturally terminates at some finite order and the relaxation function will have a compact expression. This already solves the operator dynamics for simple systems completely, see sec. \ref{finitechain} for more details. For (sufficiently) complex systems, the one dimensional chain will be semi-infinite and different approaches are needed.

\subsection{Irreversibility and ergodicity}\label{ergodic}

It turns out that the condition for the emergence of ergodic behavior for complex systems can be formulated in terms of
operator growth directly. Define a quantity $W$ as \cite{lee}
\be W=\fft{b_2^2\,b_4^2\,b_6^2\cdots}{b_1^2\,b_3^2\,b_5^2\cdots} \,,\ee
which is referred to as the canonical form of $W$. Equivalently, it can be evaluated using either the relaxation function
\be W=\phi_0(0) \,,\ee
or the auto-correlation function directly as
\be W=\int_0^{+\infty}dt\,\varphi_0(t) \,.\ee
For simple systems, since the recursion method terminates at some finite order so that $b_{2k}=0$ or $b_{2k+1}=0$, one will have $W=0$ or $W=\infty$. The process will be reversible. Otherwise, a finite $W$ (except zero) describes an irreversible process. It was shown \cite{lee} that this is equivalent to the Kubo's condition, which is formulated in terms of time averages of correlation functions. However, evaluation of $W$ provides a simpler way to test ergodicity of the theory.

 There are also other ways that were believed to probe irreversibility of operator growth. For example, at long times the auto-correlation function should approach to zero. However, as pointed out in \cite{lee}, this is only a necessary condition. In fact, it is hard to search an alternate condition equivalent to a finite $W$. Our new observation is information quantities: the relation between K-complexity and K-entropy may be such a candidate.

\section{Operator growth at initial times}

According to (\ref{newbasis}), the wave function $\varphi_n(t)$ on the one dimensional chain could be interpreted as probabilities, as in ordinary quantum mechanics. Normalization of the operator wave function $|\mc{O}(t)\rangle$ implies $\sum_{n=0}|\varphi_n(t)|^2=1$. The K-complexity was defined as the average positions of the chain \cite{Parker:2018yvk}
\be C_K:=\langle \mc{O}(t)| n |\mc{O}(t) \rangle=\sum_{n=0}n|\varphi_n|^2 \,.\ee
Clearly, it depends on the Krylov basis, the initial operator and the dynamics under considerations. Yet physical meaning of $C_K$ is far from transparent. It was established \cite{Parker:2018yvk} that K-complexity provides an upper bound for other notions of complexity, such as OTOCs. Application of the theorem to chaotic systems leads to an upper bound on Lyapunov exponent $\lambda_L\leq 2\alpha$, where $\alpha$ is the asymptotic growth rate of Lanczos coefficient $b_n\rightarrow\alpha n+\gamma$ as $n\rightarrow +\infty$. Moreover, extension of the analysis to finite temperatures, it was argued that the bound is tighter than the MSS bound \cite{Maldacena:2015waa} $\lambda_{L}\leq 2\alpha\leq 2\pi T $. Though these results are interesting, the K-complexity has not been connected to irreversibility of operator growth, as far as we know.

On the other hand, the operator entropy (or $K$-entropy) was defined as \cite{Barbon:2019wsy}
\be S_K(t):=-\sum_{n=0}|\varphi_n(t)|^2\,\mm{ln}{|\varphi_n(t)|^2} \,.\label{kentropy}\ee
Similar to $C_K$, $S_K$ depends on the Krylov basis, the initial operator and the choice of dynamics. Its physical meaning is not clear as well. Properties of $S_K$ were previously examined in the scrambling regime and the post-scrambling regime for chaotic systems \cite{Barbon:2019wsy}.

In this paper, we will investigate $C_K$ and $S_K$ for a variety of systems with dissipative behavior emerging at long times. We will show that they always enjoy a logarithmic relation $S_K\sim \log{C_K}$ in this case.

\subsection{Initial growth}
Before studying the long time dynamics, let us first address operator growth at initial times.

According to the discrete Schr\"{o}dinger equation (\ref{varphi}) and the boundary conditions, one easily finds
\be \varphi_{n}(0)=\delta_{n0}\,,\quad \dot{\varphi}_n(0)=b_1\delta_{n1}\,,\quad \ddot{\varphi}_n(0)=-b_1^2\delta_{n0}+b_1b_2\delta_{n2} \,,\ee
where a dot denotes a derivative with respect to $t$. Using these results, we deduce
\be C_K(0)=0\,,\quad \dot{C}_K(0)=0\,,\quad \ddot{C}_K(0)=2\mu_2 \,.\ee
It implies that initially, the Krylov complexity grows quadratically
\be C_K(t)=\mu_2 t^2+\cdots \,.\ee
Extension of the analysis to the operator entropy yields
\be S_K(0)=0\,,\quad \dot{S}_K(0)=0\,,\quad \ddot{S}_K(0)=-2\mu_2\,\mm{ln}{(\mu_2 t^2)}-4\mu_2 \,.\ee
To leading order, the operator entropy behaves as
\be S_K(t)=-\mu_2 t^2\,\mm{ln}{(\mu_2t^2)}+O(\mu_2t^2) \,.\ee
The results imply a product-logarithmic relation between $S_K$ and $C_K$ at initial times
\be\label{scirelation} S_K(t)=-C_K(t)\,\mm{ln}{C_K(t)}+O\big( C_K(t) \big) \,.\ee
Notice that the relation is universal to both reversible and irreversible process. It does not capture any information about long time dynamics. Comparing it to the result at long times will help us to understand the long time dynamics better.

\subsection{Testing the initial growth}

Let us test the relation (\ref{scirelation}) using several exact examples.

The first is $b_n=\alpha\sqrt{n(n-1+\eta)}$, which naturally arises in SYK model. The wave function on the semi-infinite chain can be solved as \cite{Parker:2018yvk}
\be \varphi_n(t)=\sqrt{\ft{\Gamma(n+\eta)}{n!\Gamma(\eta)}}\,\fft{\tanh^n(\alpha t)}{\cosh^{\eta}{(\alpha t)}} \,.\ee
Evaluating K-complexity yields $C_K=\sum_{n}n |\varphi_n|^2=\eta \sinh^2{(\alpha t)}$. Clearly, at initial times, $C_K\simeq \eta \alpha^2 t^2=\mu_2 t^2$, where $\mu_2=b_1^2=\eta \alpha^2$. On the other hand, $\log{|\varphi_n|^2}\simeq n \,\mm{ln}{(\alpha^2 t^2)}\simeq n \mm{ln}{C_K}$, leading to $S_K\simeq -\sum_{n}n |\varphi_n|^2 \,\mm{ln}{C_K}=-C_K\,\mm{ln}{C_K} $. This coincides with the relation (\ref{scirelation}).

 The second example is $b_n=\alpha\sqrt{n}$, which appears in several integrable models \cite{Parker:2018yvk}. The operator state $|\mc{O}(t)\rangle$ can be viewed as the Glauber coherent state in the operator Hilbert space \cite{Caputa:2021sib}. This gives
\be \varphi_n(t)=e^{-\alpha^2t^2/2}\fft{\alpha^n t^n}{\sqrt{n!}} \,.\ee
The Krylov complexity can be evaluated as $C_K=\alpha^2 t^2=\mu_2t^2$. The initially quadratic growth persists in the full time evolution. On the other hand, evaluation of the operator entropy yields
\be\label{skhw} S_K=-C_K\,\mm{ln}{C_K}+C_K+\sum_{n}|\varphi_n|^2\log{n!} \,.\ee
Clearly, at initial times, the first term on the r.h.s is dominant. This again coincides with the relation (\ref{scirelation}).

The third example is $b_n=\alpha\sqrt{n(2j-n+1)}$, where $j$ is an integer or half integer \cite{Caputa:2021sib}. The operator Hilbert space has a finite dimension with $0\leq n\leq 2j$. It describes a reversible process since $W=\infty$ ( or $W=0$) for an integer (or half integer) $j$. One has
\be \varphi_n(t)=\sqrt{\ft{\Gamma(2j+1)}{n!\Gamma(2j-n+1)}}\,\fft{\tan^n(\alpha t)}{\cos^{-2j}{(\alpha t)}} \,.\ee
 Evaluation of K-complexity and K-entropy yields $C_K=2j\sin^2{(\alpha  t)}$ and
\be S_K=-C_K\,\mm{ln}{\tan^2{(\alpha t)}}-\,\mm{ln}{\Big( \Gamma(2j+1)\cos^{4j}(\alpha t)\Big)}+\sum_{n}|\varphi_n|^2\,\mm{ln}{\Big( n!\,\Gamma(2j-n+1) \Big)} \,.\ee
Again at initial times $C_K\simeq \mu_2 t^2$ and $S_K\simeq -C_K\,\mm{ln}{C_K}$, consistent with the relation (\ref{scirelation}).

In fact, for reversible process, the operator wave functions can always be solved exactly and hence there are a lot of examples that can test the relation (\ref{scirelation}), for example see (\ref{finite1}).

\section{Operator growth at long times}

We are moving to study the functional relation between K-complexity and K-entropy for a variety of dissipative systems, including chaotic ones, integrable theories and many others.

We will adopt the continuum limit as well as numerical approach. For semi-infinite chains, the continuum limit is good at capture long time behaviors of K-complexity and K-entropy using coarse grained wave functions. We shall briefly review it by following \cite{Barbon:2019wsy}.

  Introducing a lattice cutoff $\epsilon$ and defining a coordinate $x=\epsilon n$ and velocity $v(x)=2\epsilon b_n$. The interpolating wave function is defined as $\varphi(x\,,t)=\varphi_n(t)$. Continuum version of the discrete equation (\ref{varphi}) is given by
\be \partial_t\varphi(x\,,t)=\fft{1}{2\epsilon}\Big[ v(x)\varphi(x-\epsilon)-v(x+\epsilon)\varphi(x+\epsilon) \Big] \,. \ee
Expansion in powers of $\epsilon$, one finds a chiral wave equation to leading order
\be\label{expansion} \partial_t\varphi=-v(x)\partial_x\varphi-\fft12 \partial_x v(x)\varphi+O(\epsilon) \,,\ee
with a position-dependent velocity $v(x)$ and mass $\fft12 \partial_x v(x)$. Introducing a new coordinate $y$ as $v(x)\partial_x=\partial_y$ and a rescaled wave function
\be \psi(y\,,t)=\sqrt{v(y)}\,\varphi(y\,,t) \,,\ee
the equation simplifies to
\be\label{chiralwave} (\partial_t+\partial_y)\psi(y\,,t)=0+O(\epsilon) \,.\ee
The general solution is given by
\be \psi(y\,,t)=\psi_i(y-t) \,,\ee
where $\psi_i(y)=\psi(y\,,0)$ stands for the initial amplitude. It implies that at leading order the coarse grained wave function moves ballistically. It turns out that this leading order approximation always derives the growth of K-complexity correctly. However, for K-entropy, some higher order corrections should be included for certain cases. In fact, the method just captures the leading long time dependence for both K-complexity and K-entropy qualitatively. Hence, we will also adopt numerical approach as a supplement.

Normalization condition in both $x$-frame and $y$-frame reads
\be 1=\sum_{n}|\varphi_n(t)|^2=\fft{1}{\epsilon}\int \mm{d}x\, \varphi^2(x\,,t)=\fft{1}{\epsilon}\int \mm{d}y\, \psi^2(y\,,t)  \,.\ee
The K-complexity can be evaluated as
\bea\label{geneck}
C_K(t)&=&\sum_{n}n |\varphi_n(t)|^2=\fft{1}{\epsilon}\int\mm{d}x\,\fft{x}{\epsilon}\,\varphi^2(x\,,t) \nn\\
      &=&\fft{1}{\epsilon}\int\mm{d}y\, \fft{x(y)}{\epsilon}\, \psi_i^2(y-t) \nn\\
      &=&\fft{1}{\epsilon}\int\mm{d}y\, \fft{x(y+t)}{\epsilon}\,\, \psi_i^2(y) \,,\eea
and the K-entropy reads
\bea\label{genesk}
S_K(t)&=&-\sum_{n}|\varphi_n(t)|^2\,\mm{ln}{|\varphi_n(t)|^2} \nn\\
      &=&-\fft{1}{\epsilon}\int\mm{d}x\,\varphi^2(x\,,t)\,\mm{ln}{\varphi^2(x\,,t)} \nn\\
      &=&-\fft{1}{\epsilon}\int\mm{d}y\, \psi_i^2(y-t)\,\mm{ln}{\Big[\fft{\psi_i^2(y-t)}{v(y)} \Big]} \nn\\
      &=&-\fft{1}{\epsilon}\int\mm{d}y\, \psi_i^2(y)\,\mm{ln}{\psi_i^2(y)}+\fft{1}{\epsilon}\int\mm{d}y\, \psi_i^2(y)\,\mm{ln}{v(y+t)} \,,\eea
where the time dependence of $S_K$ is only contained in the second term on the r.h.s. Once we know the transformation between the two frames, we are able to extract the leading time dependence of $C_K$ and $S_K$ at long times immediately.

\subsection{Chaotic systems}

It was first conjectured in \cite{Parker:2018yvk} that for chaotic systems, the Lanczos coefficient grows asymptotically linearly $b_n\rightarrow\alpha n+\gamma$ as $n\rightarrow +\infty$ (This is valid to the infinite temperature limit. At finite temperatures, it was shown in \cite{Dymarsky:2021bjq} that the asymptotically linear behavior of $b_n$ can also be obtained for free quantum field theories, which however do not probe chaos ). This gives rise to $v(x)=2\alpha x+2\epsilon\gamma$, where the subleading order correction $\gamma$ plays only a short time in the scrambling regime and hence is negligible in the continuum limit. One finds
\be y=\fft{1}{2\alpha}\mm{ln}{\big( \fft{x}{\epsilon} \big)} \qquad \mm{or}\qquad x=\epsilon\, e^{2\alpha y} \,.\ee
Evaluation of K-complexity yields
\be C_K(t)=\fft{1}{\epsilon}\int\mm{d}y\, e^{2\alpha(y+t)}\, \psi_i^2(y)=C_K(0)\,e^{2\alpha t} \,.\ee
It grows exponentially with correct exponent. This coincides with the SYK model. The operator entropy is given by
\bea S_K(t)&=&\fft{1}{\epsilon}\int\mm{d}y\, \psi_i^2(y)\,\mm{ln}{v(y+t)}+\cdots  \nn\\
&=&\fft{1}{\epsilon}\int\mm{d}y\, \psi_i^2(y)\,2\alpha(y+t)+\cdots \nn\\
&=&2\alpha t+\cdots \,,\eea
which grows linearly with time. These results have already been derived in \cite{Barbon:2019wsy}. Our new contribution here is we realize that at long times\footnote{Asymptotic growth of $C_K$ implies that the time scale at which the logarithmic relation emerges may be defined as $C_K\sim O(1)$. This coincides with the scrambling time $\sim \log S$ for chaotic systems, where $S$ stands for the number of degrees of freedoms.} they imply a logarithmic relation
\be\label{skchaotic0} S_K(t)=\mm{ln}{C_K(t)}+\cdots \,,\ee
 which is very similar to the celebrated Boltzmann relation in statistical mechanics. Indeed, operator randomization is very efficient for fast scramblers \cite{Barbon:2019wsy} so that a statistical description should emerge in the scrambling regime (and beyond). This inspires us that the above relation may signal irreversibility of operator growth for general cases, irrespective of choice of dynamics. This motivates us to study the relation between $S_K$ and $C_K$ for many other systems.

However, it should be emphasized that in the relation (\ref{skchaotic0}) the proportional coefficient is undetermined since continuum limit just captures the leading time dependence of $C_K$ and $S_K$ qualitatively. In general, we may take the relation in the form of Eq.(\ref{skchaotic}). However, it turns out that the relation (\ref{skchaotic0}) is correct for chaotic systems. For example, consider the SYK model with $\eta=1$, the K-complexity and K-entropy can be evaluated exactly as
\bea
&&C_K=\sinh^2{(\alpha t)}\,,\nn\\
&& S_K=\cosh^2{(\alpha t)}\,\mm{ln}{\cosh^2{(\alpha t)}}-\sinh^2{(\alpha t)}\,\mm{ln}{\sinh^2{(\alpha t)}}\,.
\eea
In the long time limit, $C_K\rightarrow e^{2\alpha t}/4$ and $S_K\rightarrow 2\alpha t$, which leads to the relation (\ref{skchaotic0}), with the proportional coefficient exactly equals to unity. For generic $\eta$, we find that this is always true. From a statistical point of view, the Boltzmann-like relation (\ref{skchaotic0}) emerges from a uniform distribution. This inspires us that the relation may hold for general chaotic systems, in which operator randomization is most efficient. We check this idea for a variety of chaotic models and find that it is indeed true. It also implies that for chaotic systems, the operator wave functions at long times are effectively described by a uniform distribution at leading order.

As an example, consider a class of model spectral density \cite{vsmuller}
\be\label{modelspectral} \Phi(\omega)=\fft{\pi}{\omega_0 \Gamma(\nu+1)} \Big| \fft{\omega}{\omega_0} \Big|^\nu \mm{exp}\Big( -\big| \fft{\omega}{\omega_0} \big| \Big) \,,\ee
where $\omega_0=2\alpha/\pi$. Notice that exponential decay of $\Phi(\omega)$ at large frequency is equivalent to asymptotically linear growth of Lanczos coefficient \cite{vsmuller}. Hence, the model spectral density describes certain chaotic systems. It turns out that in this case the frequency moments can be written in closed form as
\be \mu_{2n}=\omega_0^{2n}\,\Gamma(1+\nu+2n)/\Gamma(1+\nu) \,.\ee
The Lanczos coefficients can be computed using recurrence relations \cite{vsmuller}. The auto-correlation function turns out to be
\be C(t)=\varphi_0(t)=\fft{1}{2\big(1-i\omega_0 t \big)^{1+\nu}}+\fft{1}{2\big(1+i\omega_0 t \big)^{1+\nu}} \,.\ee
\begin{figure}\label{fig1}
\centering
  \includegraphics[width=250pt]{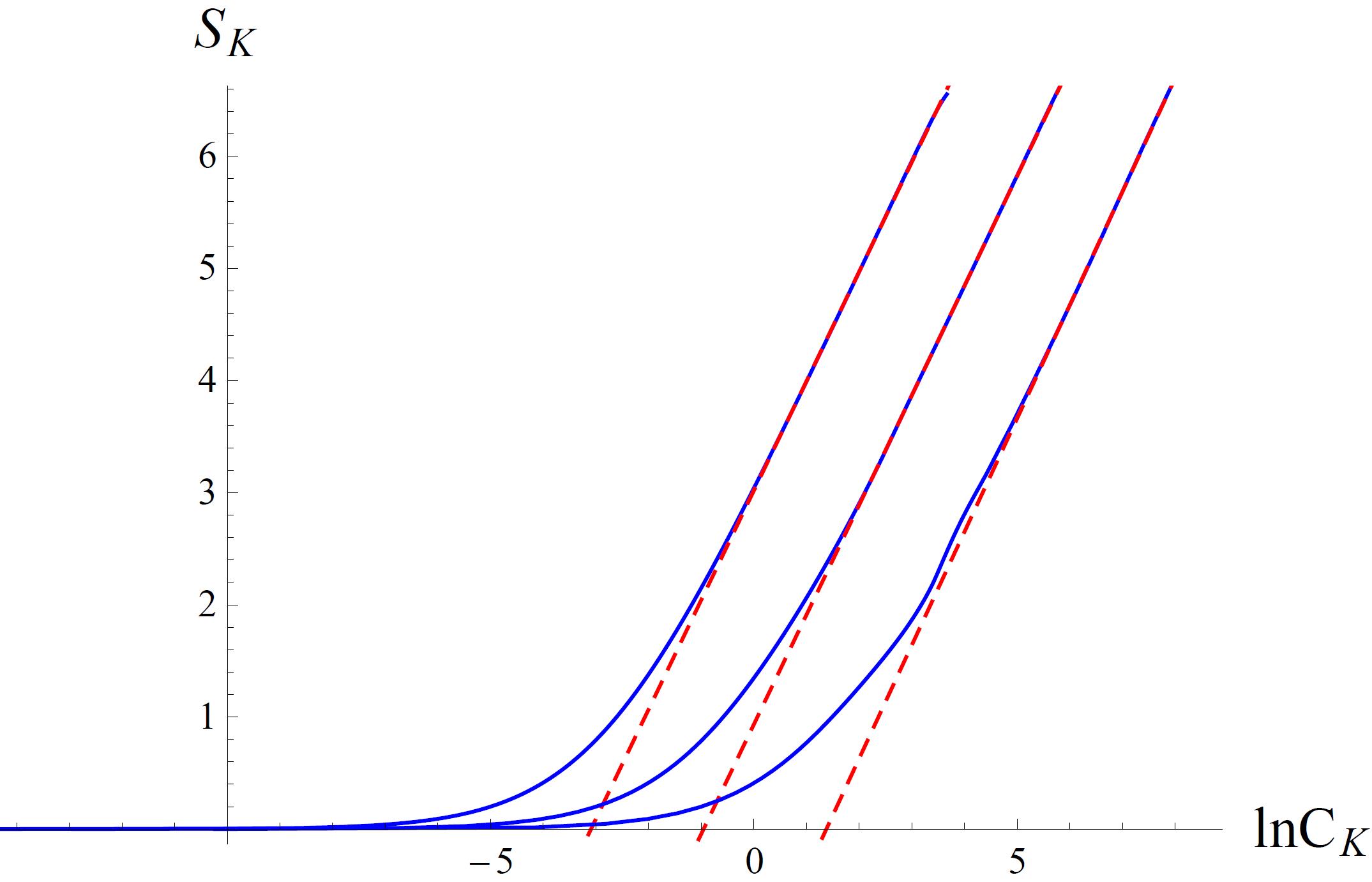}
  \caption{The functional relation $S_K\sim \mm{ln} C_K$ is shown for the model spectral density (\ref{modelspectral}), where $\nu=0\,,1\,,2$ from left to right. At long times, the linear relation Eq.(1) holds, with (shown in dashed lines) $S_K=0.976348\,\mm{ln}{C_K}+1.06661$ ($\nu=0$), $S_K=0.978129\,\mm{ln}{C_K}+0.934119$ ($\nu=1$) and $S_K=1.01102\,\mm{ln}{C_K}+0.63022$ ($\nu=2$). Within our numerical accuracy $\tilde{\eta}\simeq 1$ for all these cases. To have a nice presentation, we have moved the results along the horizontal axis properly.}
\end{figure}
The remaining wave functions $\varphi_n(t)$'s can be deduced using the discrete equation (\ref{varphi}). To simplify matter, we set $\omega_0=2/\pi$ so that $\alpha=1$. The functional relation $S_K(C_K)$ for several $\nu$'s value was shown numerically in Fig. 1. It is easily seen that at long times the logarithmic relation (\ref{skchaotic}) indeed holds, with the proportional coefficient $\tilde{\eta}$ close to unity\footnote{Because of exponential growth of $C_K$, computational cost increases exponentially for chaotic models. This limits our numerical accuracy.}
\bea
&&\nu=0\,,\qquad \tilde{\eta}=0.976348\,,\nn\\
&&\nu=1\,,\qquad \tilde{\eta}=0.978129\,,\nn\\
&&\nu=2\,,\qquad \tilde{\eta}=1.011020\,.
\eea
This supports our intuitive idea that the Boltzmann relation (\ref{skchaotic0}) holds for general chaotic systems.

To end this subsection, let us comment on the case beyond the scrambling regime. It was argued \cite{Barbon:2019wsy} that in this case, the Lanczos coefficient will approach to a constant $b_n\rightarrow b=\alpha S$, for systems with $S$ extensive degrees of freedoms. Using continuum limit, it was shown that K-complexity grows linearly
\be C_K(t)_{\mm{post-scrambling}}\sim 2b(t-t_*)+\cdots\,, \ee
until arriving at its maximum value of order $\sim e^{O(S)}$, where $t_*$ stands for the scrambling time. On the other hand, a careful examination of the long time tails of wave functions leads to
\be S_K(t)\sim \mm{ln}{\big( 2bt \big)}+\cdots\,. \ee
The operator entropy continues to grow until arriving at the maximum value $S_K\sim O(S)$. This again leads to the logarithmic relation (\ref{skchaotic}), though the constant $\tilde\eta$ is undetermined. However, we argue that in this case $\tilde\eta$ should be still equal to unity since in the post-scrambling regime, the operator wave functions should still obey a uniform distribution to leading order.

The above result supports our intuitive idea that the logarithmic relation (\ref{skchaotic}) simply signals irreversibility, irrespective of dynamical details of the process. We would like to extend the analysis to general systems with emergence of dissipative behaviors, where the proportional constant $\tilde{\eta}$ is undetermined. Here it is worth emphasizing that the K-entropy approaches to its maximum value for a uniform distribution because of maximal entropy principle. Hence, we propose that the constant $\tilde{\eta}$ is bounded as
\be 0<\tilde{\eta}\leq 1 \,,\ee
where the up bound is saturated for fast scramblers, including $\mm{1d}$ chaotic systems. We find that this is true as far as we can check.

\subsubsection{With logarithmic correction}

It was argued in \cite{Parker:2018yvk} that for $\mm{1d}$ chaotic systems, the asymptotic behavior of Lanczos coefficient acquires a logarithmic correction: $b_n=\alpha n/\mm{ln}{n}+\cdots$. However, for our purpose, we may take it as a nearly chaotic model in diverse dimensions: the Lanczos coefficient grows faster than integrable theories (which have a power law $b_n\sim n^\delta\,,0<\delta<1$) but still slower than fast scramblers. A more general case may be taken as $b_n=\alpha n/(\mm{ln}{n})^{\sigma}$, where $\sigma>0$, corresponding to a velocity $v(x)=2\alpha x/(\mm{ln}{\ft{x}{\epsilon}})^\sigma$. In this case, the two frames are connected by
\begin{figure}
  \centering
  \includegraphics[width=210pt]{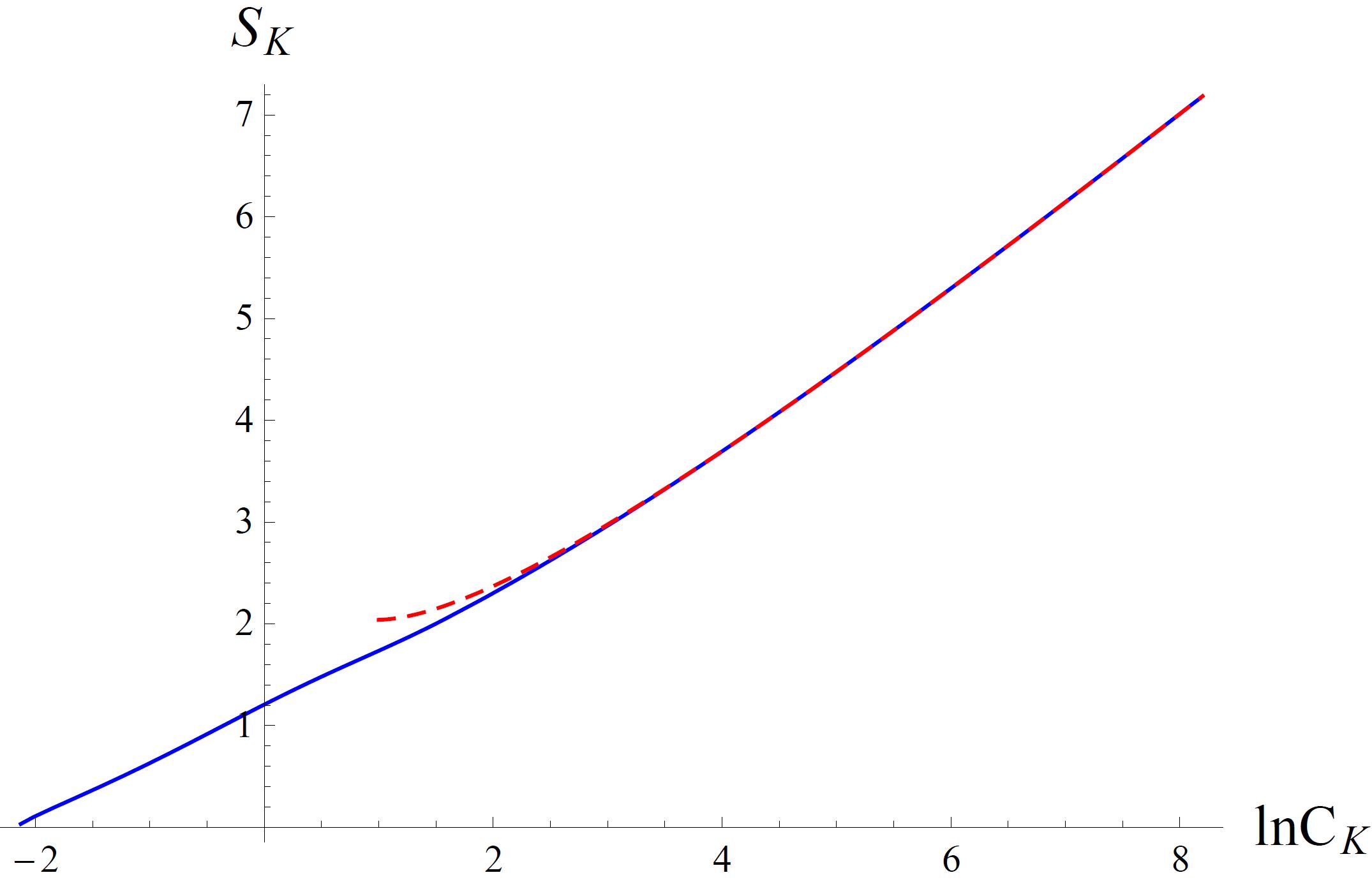}
  \includegraphics[width=210pt]{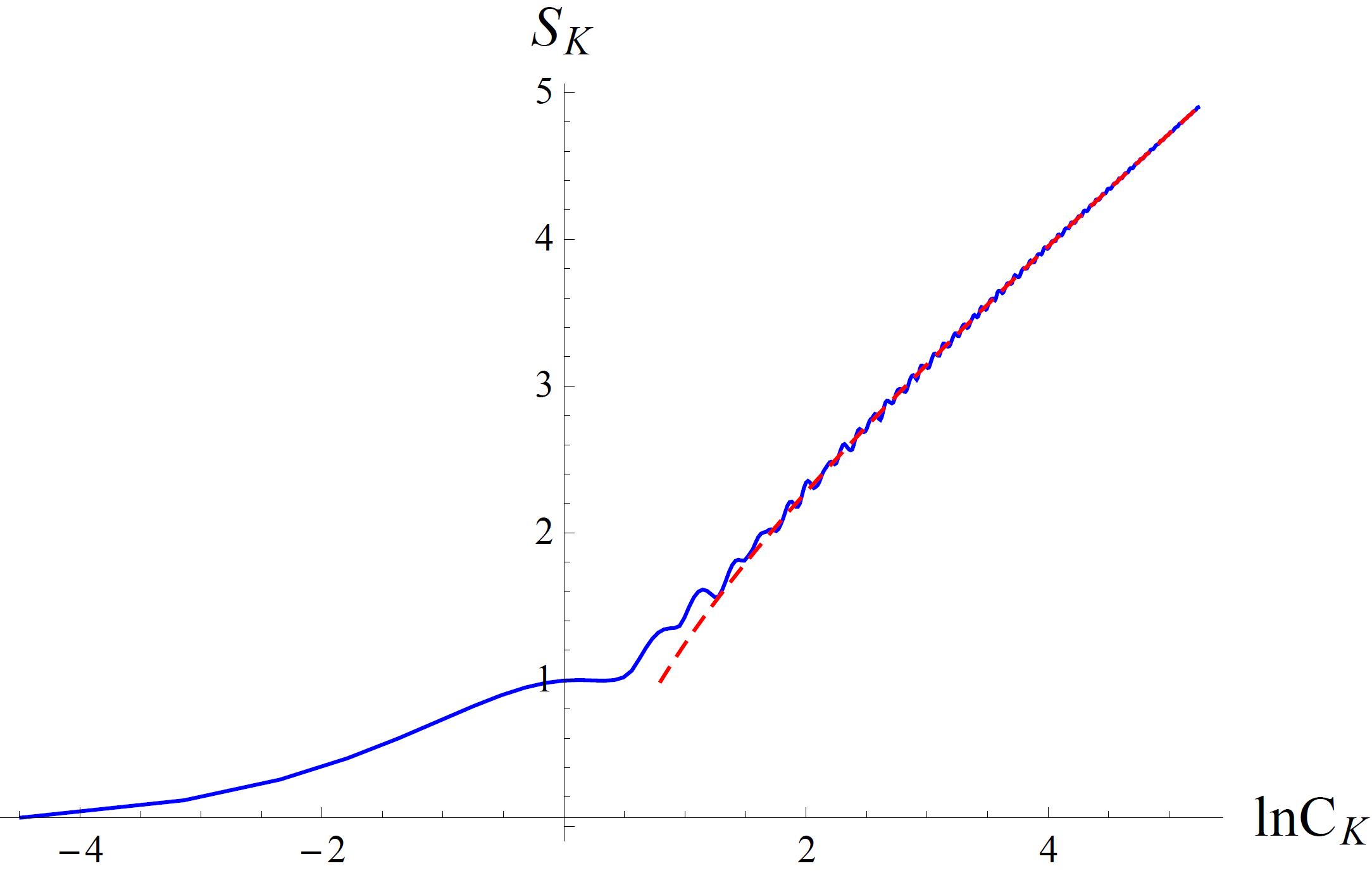}
  \caption{The functional relation $S_K(C_K)$ for nearly chaotic models. $b_n=\ft{\alpha n}{\mm{ln}{(n+1)}}$ for the left panel and $b_n=\ft{\alpha n}{\mm{ln}^2{(n+1)}}$ for the right panel. We set $\alpha=1$ in numerical calculations. At long times, the functional relation is very well fitted by $S_K=0.994787\,\mm{ln}{C_K}-0.958367\,\mm{ln}{\mm{ln}{C_K}}+1.04432$ (left) and $S_K=0.634549\,\mm{ln}{C_K}+0.593231\,\mm{ln}{\mm{ln}{C_K}}+0.59138$ (right), matching the continuum limit. }
  \label{quasichaotic}
\end{figure}
\be x=\epsilon\,\mm{exp}\Big[ \big(2\tilde{\alpha}y \big)^{\fft{1}{1+\sigma}}\Big] \,,\ee
where $\tilde{\alpha}=\alpha(1+\sigma)$. It is straightforward to deduce
\be
C_K(t)=\fft{1}{\epsilon}\int \mm{d}y\,\psi_i^2(y)\,\mm{exp}\Big[ \big(2\tilde{\alpha}(y+t) \big)^{\fft{1}{1+\sigma}}\Big]\sim \mm{exp}\Big[ \big(2\tilde{\alpha}t \big)^{\fft{1}{1+\sigma}}\Big]\,,
\ee
and
\bea S_K(t)&=&\fft{1}{\epsilon}\int \mm{d}y\,\psi_i^2(y)\,\Big[\big(2\tilde{\alpha}(y+t) \big)^{\fft{1}{1+\sigma}}-\mm{ln}{\big(2\tilde{\alpha}(y+t) \big)^{\fft{1}{1+\sigma}}} \Big]+\cdots\nn\\
&\sim& \big(2\tilde{\alpha} t \big)^{\fft{1}{1+\sigma}}-\mm{ln}{\big(2\tilde{\alpha}t \big)^{\fft{1}{1+\sigma}}}+\cdots  \,,\eea
where a tilde means taking the long time limit and we have ignored the constant coefficient in each term. Combining these results again leads to the logarithmic relation (\ref{skchaotic}) except that the subleading order correction is of order $O(\mm{ln}\mm{ln}{C_K})$. However, from the method itself, we do not expect it can extract the subleading order term of $S_K$ (or $C_K$) correctly since the leading term is only determined qualitatively. Nevertheless, the analysis is consistent with our numerical calculations for several examples, see Fig. \ref{quasichaotic}. However, numerically it is also hard to ensure the form of the subleading order corrections. For numerical data established in the figure, including $\mm{ln}\mm{ln}{C_K}$ term just fits the numerical data slightly better than that without it but it should not be considered conclusive. However, an exception occurs for $\mm{1d}$ chaotic systems. We find that in this case $\tilde\eta$ is equal to unity if and only if the $\mm{ln}\mm{ln}{C_K}$ term is included!


\subsection{Integrable theories}
Theories with asymptotic growth of Lanczos cofficients $b_n\rightarrow\alpha n^\delta$, where $0<\delta<1$ were referred to as integrable ones \cite{Parker:2018yvk}. In this case, $v(x)=2\alpha \epsilon^{1-\delta}x^\delta$ and the two frames are connected by
\be y=\fft{1}{2\tilde{\alpha}}\Big( \fft{x}{\epsilon} \Big)^{1-\delta}\qquad \mm{or}\qquad x=\epsilon\, \big(2\tilde{\alpha}y \big)^{\fft{1}{1-\delta}} \,,\ee
where $\tilde{\alpha}=\alpha(1-\delta)$.
Evaluation of K-complexity yields
\be
C_K(t)=\fft{1}{\epsilon}\int\mm{d}y\, \big[2\tilde{\alpha}(y+t) \big]^{\fft{1}{1-\delta}}\, \psi_i^2(y)\sim \big( 2\tilde{\alpha} t\big)^{\fft{1}{1-\delta}}+\cdots \,.\ee
It grows in a power law at sufficiently long time. On the other hand, the operator entropy is given by
\bea
S_K(t)&=&\fft{1}{\epsilon}\int\mm{d}y\,\psi_i^2(y)\,\mm{ln}{\big[2\tilde{\alpha}(y+t)\big]^\fft{\delta}{1-\delta}}+\cdots  \nn\\
      & \sim &\fft{\delta}{1-\delta}\,\mm{ln}{\big( 2\tilde{\alpha} t\big)}+\cdots   \nn\\
      & \sim &\delta \,\mm{ln}{C_K(t)}+\cdots \,.
 \eea
Again this leads to the logarithmic relation (\ref{skchaotic}). However, unlike previous cases, the proportional constant $\tilde \eta=\delta$ is exact for $\delta\geq 1/2$ (from the method itself, this should be simply a coincidence). We check it using a lot of numerical examples and find that it is always true. For example, consider integrable models with $b_n=\alpha n^{\delta}$. Numerically evaluation of the operator entropy (\ref{skhw}) yields at long times
\bea
&&\delta=1/2\,,\qquad S_K(t)= 0.500917\,\mm{ln}{C_K(t)}+1.41373\,,\nn\\
&&\delta=2/3\,,\qquad S_K(t)= 0.669477\,\mm{ln}{C_K(t)}+1.43501\,,\nn\\
&&\delta=3/4\,,\qquad S_K(t)= 0.752683\,\mm{ln}{C_K(t)}+1.44995\,.
\eea
In all these cases, the proportional constant $\tilde\eta$ is equal to $\delta$, within our numerical accuracy. It is a pity that we do not have a physical interpretation for this.

\subsubsection{With logarithmic correction}

To test whether the above result can be extended to slightly different cases, let us consider logarithmic corrections to integrable models. For instance, the Lanczos coefficient behaves asymptotically as $b_n=\alpha n^\delta/\mm{ln}{n}$, giving rise to $v(x)=2\alpha\epsilon \big( \ft{x}{\epsilon}\big)^\delta/\mm{ln}{\big( \ft{x}{\epsilon} \big)}$. The two frames are connected by
\be y=\fft{1}{2\bar{\alpha}}\big( X \,\mm{ln}{X}-X\big) \,,\quad X=\big(\ft{x}{\epsilon} \big)^{1-\delta}\,,\ee
where $\bar\alpha=\alpha(1-\delta)^2$. Notice at very large $y$, $X\sim (2\bar{\alpha}y)/\mm{ln}{(2\bar{\alpha}y)}$ to leading order. Hence, at long times the K-complexity and K-entropy behave as
\bea
&& C_K(t)\sim x(t)/\epsilon\sim (2\bar{\alpha}t)^\gamma/\mm{ln}^\gamma{(2\bar{\alpha}t)}\,,\nn\\
&& S_K(t)\sim \mm{ln}{v(t)}\sim \delta \,\mm{ln}\big( \ft{x(t)}{\epsilon}\big)\sim \delta \,\mm{ln}{C_K(t)}\,,
\eea
where $\gamma=1/(1-\delta)$. It turns out that the K-complexity grows a bit slower than the integrable models. However, while the logarithmic relation (\ref{skchaotic}) still holds, the proportional coefficient $\tilde\eta=\delta$ is falsified by our numerical results.

One may also consider positively corrections to Lanzos coefficient as $b_n=\alpha n^\delta \,\mm{ln}{n}$, corresponding to $v(x)=2\alpha\epsilon \big( \ft{x}{\epsilon}\big)^\delta \,\mm{ln}{\big( \ft{x}{\epsilon} \big)}$. The two frames are connected as
\be y=\ft{1}{2\alpha}\mm{li}(X) \,,\ee
where $\mm{li}(x)$ stands for the logarithmic integral function. At large $y$, one has $X\sim 2\alpha y\,\mm{ln}{(2\alpha y)}$, implying that the K-complexity grows a bit faster than integrable models
\be C_K\sim  (2\alpha t)^\gamma \,\mm{ln}^\gamma{(2\alpha t)}\,.\ee
Again the K-entropy turns out to be $S_K(t)\sim \mm{ln}{v(t)}\sim \delta\,\mm{ln}{C_K(t)}$, but the constant coefficient $\tilde\eta=\delta$ is falsified by our numerical results. It seems that the result $\tilde\eta=\delta$ is only valid to integrable theories with $\delta\geq 1/2$.


\subsection{With bounded support and beyond}
Consider $\delta\rightarrow 0$ limit of integrable theories. The Lanczos coefficient approaches to a constant $b_n\rightarrow b$ asymptotically, corresponding to a constant velocity $v(x)=v=2\epsilon b$. This case is very similar to the post-scrambling regimes of chaotic systems \cite{Barbon:2019wsy}, except the initial amplitude. Careful analysis using the continuum limit (with higher order corrections) implies that the K-entropy grows logarithmically at long times
\be S_K(t)\sim \mm{ln}{(2bt)}\,, \ee
whereas the K-complexity increases linearly $C_K(t)\sim 2bt$. This again leads to the logarithmic relation (\ref{skchaotic}).

To test the result, consider two simple models\footnote{These two models arise from a chain of classical harmonic oscillator of $2N$ atoms with periodic boundary conditions \cite{Lee1989}. }. The first is $b_n=\omega_0/2$. The wave function is solved in terms of Bessel functions
\be \varphi_n(t)=J_n(\omega_0 t)+J_{n+2}(\omega_0t) \,.\ee
Numerically evaluation of K-complexity and K-entropy yields at long times
\be S_K(t)=0.729302\,\mm{ln}{C_K(t)}+0.353124 \,.\ee
The second example is $b_1=\omega_0/\sqrt{2}\,,b_n=\omega_0/2$ for $n>1$. The wave function is given by
\be \varphi_n(t)=c_n J_n(\omega_0t) \,, \ee
where $c_{-1}=0\,,c_0=1$ and $c_n=\sqrt{2}$ for $n>1$. We find at long times
\be S_K(t)=0.870024 \,\mm{ln}{C_K(t)}+0.603974  \,.\ee
These results support the continuum limit analysis.

To proceed, consider a critical case where the Lanczos coefficient grows asymptotically faster than the bounded case but still slower than a power law. For example: $b_n=\alpha \,\mm{ln}{n}$. It corresponds to a velocity $v(x)=2\epsilon\alpha \,\mm{ln}{(\ft{x}{\epsilon})} $ and the two frames are transformed as
\be y(x)=\ft{1}{2\alpha}\mm{li}\big(\ft{x}{\epsilon} \big) \,.\ee
Since at large $y$, $x/\epsilon\sim 2\alpha y\,\mm{ln}{(2\alpha y)}$, one has
\be C_K(t)\sim x(t)/\epsilon\sim 2\alpha t\,\mm{ln}{(2\alpha t)}\,,\ee
 which is a product logarithmic law. The K-complexity grows faster than the bounded case (which is linear) but still slower than the integrable theories. This is in accordance with our expectations.
\begin{figure}
  \centering
  \includegraphics[width=210pt]{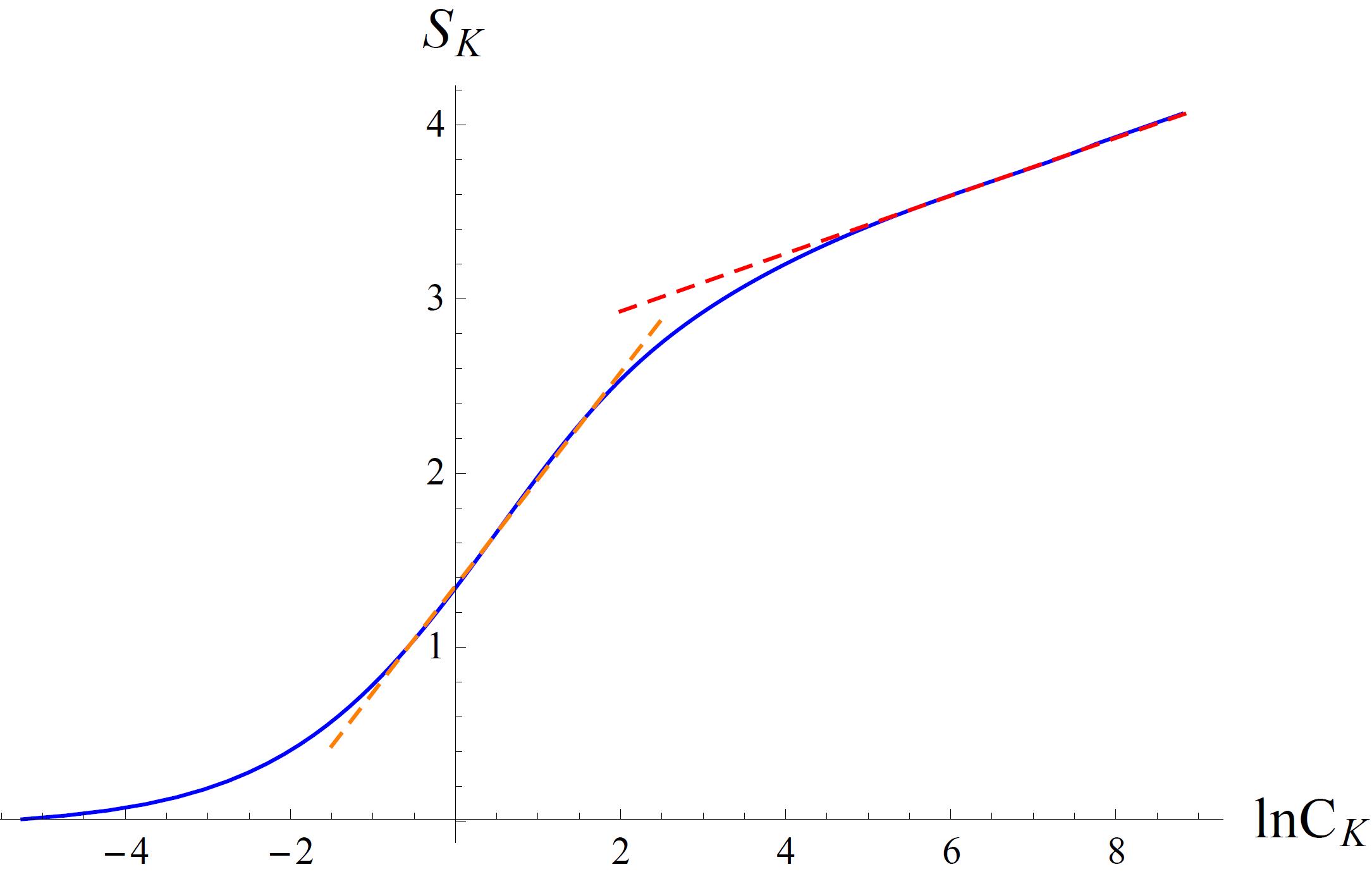}
  \includegraphics[width=210pt]{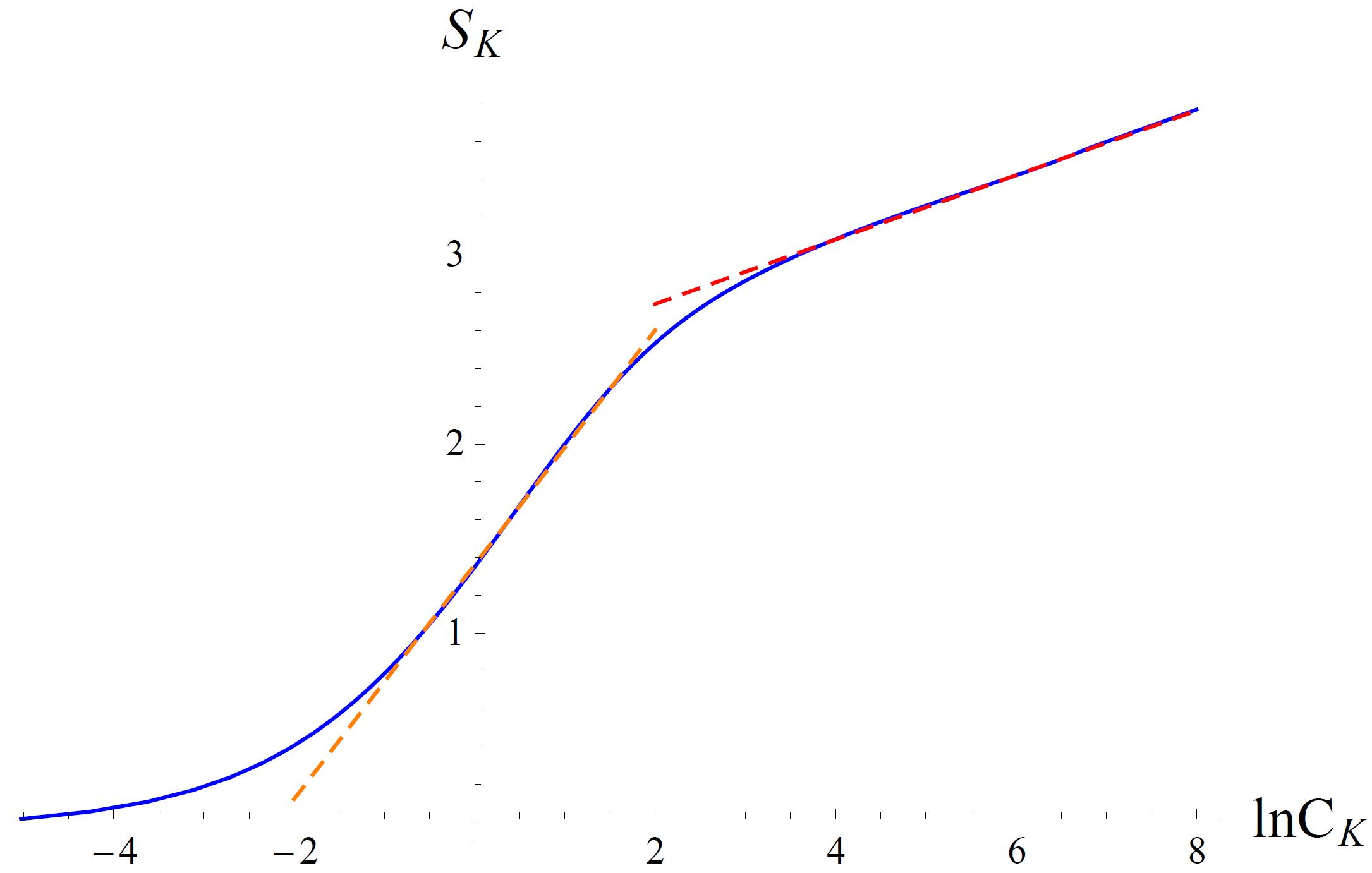}
  \caption{Left panel: $b_n=\alpha\,\mm{ln}{(n+1)}$. Right panel: $b_n=\alpha\,\mm{ln}{n}+\gamma$. We set $\alpha=\gamma=1$. In both cases, the logarithmic relation (\ref{skchaotic}) already appears at the time scale where $C_K(t_c)\sim O(1) $ (see the orange lines) but the coefficient $\tilde\eta$ is changed at later times (see the red lines). Around $t\sim t_c$, we have $S_K=0.613581\,\mm{ln}{C_K}+1.35088$ (left) and $S_K=0.619693\,\mm{ln}{C_K}+1.36223$ (right) whereas at long times $S_K=0.166437\,\mm{ln}{C_K}+2.59585$ (left) and $S_K=0.170329\,\mm{ln}{C_K}+2.40017$ (right). Within our numerical accuracy, both cases have the same coefficient $\tilde\eta$ at the same time regimes.}
  \label{quasibound}
\end{figure}

However, at this order, the K-entropy turns out to be $S_K\sim \mm{ln} v(t)\sim \mm{ln}\mm{ln}{C_K}$, violating the logarithmic relation (\ref{skchaotic}) apparently. To resolve the issue, one may include higher order corrections in the wave equation as the bounded case. Unfortunately, we do not find a definite answer using this approach. Nevertheless, our numerical results still suggest that the logarithmic relation (\ref{skchaotic}) holds in this case, see Fig. \ref{quasibound}.

To end this section, let us discuss the relation $S_K(C_K)$ in the full time evolution. We have seen that for systems described by semi-infinite chains $S_K\sim -C_K\,\mm{ln}{C_K}$ at initial times and $S_K\sim \mm{ln}{C_K}$ at long times where dissipative behavior emerges. These two regimes are universal, irrespective of the choice of dynamics. The former is determined by the lowest moment $\mu_2$ whereas the latter probably signals irreversibility of the process. There is a smooth crossover between them, which however depends on the dynamical details. Since the functional function $S_K(C_K)$
is well defined in the full time evolution, it makes sense to study it in operator dynamics and connect it to emergence of ergodic behavior of the theories.


\section{Finite chains}\label{finitechain}
In this section, we would like to study K-complexity and K-entropy for finite chains and compare the results with previous cases. In this case, the recursion method naturally comes to a stop at some order $K$ so that $b_{K+1}=0$. As a consequence, the continued-fraction representation of the relaxation function terminates at the $K$-th level:
\be
\phi_0(z)=\dfrac{1}{z+\dfrac{b_1^2}{z+\dfrac{b_2^2}{  \begin{array}{cc}z+\cdots &  \\  &\cdots+\dfrac{b_K^2}{z} \end{array}  } } } \,.
\ee
This is a rational function $p_K(z)/q_{K+1}(z)$. Here there are two cases, depending on $K$ is odd or even, corresponding to $W=0$ or $W=\infty$. For odd $K$,
\bea\label{pq}
&&p_K(z)=z^K+ z^{K-2}+\cdots+ z \,,\nn\\
&&q_{K+1}(z)=z^{K+1}+ z^{K-1}+\cdots+c\,,
\eea
where we have omitted the constant coefficient for each term in the polynomials and $c\neq 0$. This leads to $W=p_K(0)/q_{K+1}(0)=0$. On the other hand, for even $K$, the results (\ref{pq}) still hold but with the last term interchanged between $p_K(z)$ and $q_{K+1}(z)$, giving rise to $W=\infty$. Hence, for both cases, the dynamics of operator is nonergodic. We will show that as a consequence, the functional relation $S_K\sim \log{C_K}$ at long times does not hold any longer.

To proceed, consider odd $K$ at first. The relaxation function contains $K+1$ poles, which are all located on the imaginary axis. The spectral function, inferred via
\be \Phi(\omega)=\lim_{\varepsilon\rightarrow 0}2\mm{Re}\big[ \phi_0(\varepsilon-i\omega)\big]  \,,\ee
consists of $L=(K+1)/2$ pairs of $\delta$-functions
\be \Phi(\omega)=\pi \sum_{\ell=1}^L a_\ell\big[\delta(\omega-\omega_\ell)+\delta(\omega+\omega_\ell)\big] \,,\ee
where all the frequencies $\omega_\ell$ are generally nonzero. The auto-correlation function turns out to be
\be\label{finitephi0} \varphi_0(t)=\sum_{\ell=1}^{L}a_\ell \cos{(\omega_\ell t)} \,.\ee
However, if one of the $L$ frequencies $\omega_\ell$ happens to be zero, the total number of Lanczos coefficients will be reduced by one so that $K=2L-2$. This is exactly the even $K$ case. Nevertheless, the result (\ref{finitephi0}) is valid to both cases.

 Given the above auto-correlation function, it turns out that all the remaining (nonzero) wave functions $\varphi_n$'s will be a sum of sine or cosine functions. It is immediately seen that in this case both K-complexity and K-entropy will no longer grow monotonically in the time evolution. Of course, the functional relation $S_K\sim \mm{ln}{C_K}$ will not hold at long times any longer.
\begin{figure}
  \includegraphics[width=140pt]{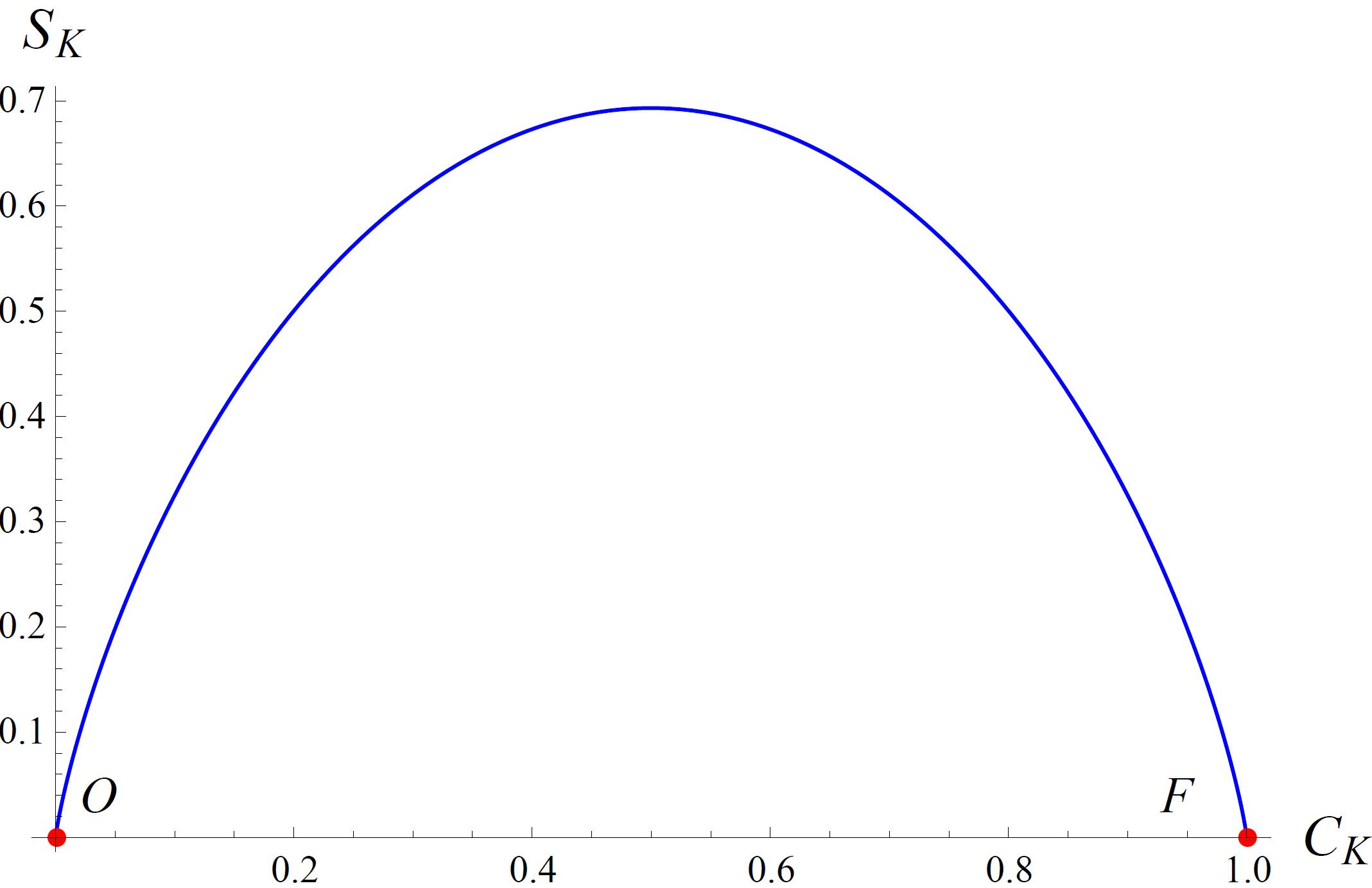}
  \includegraphics[width=140pt]{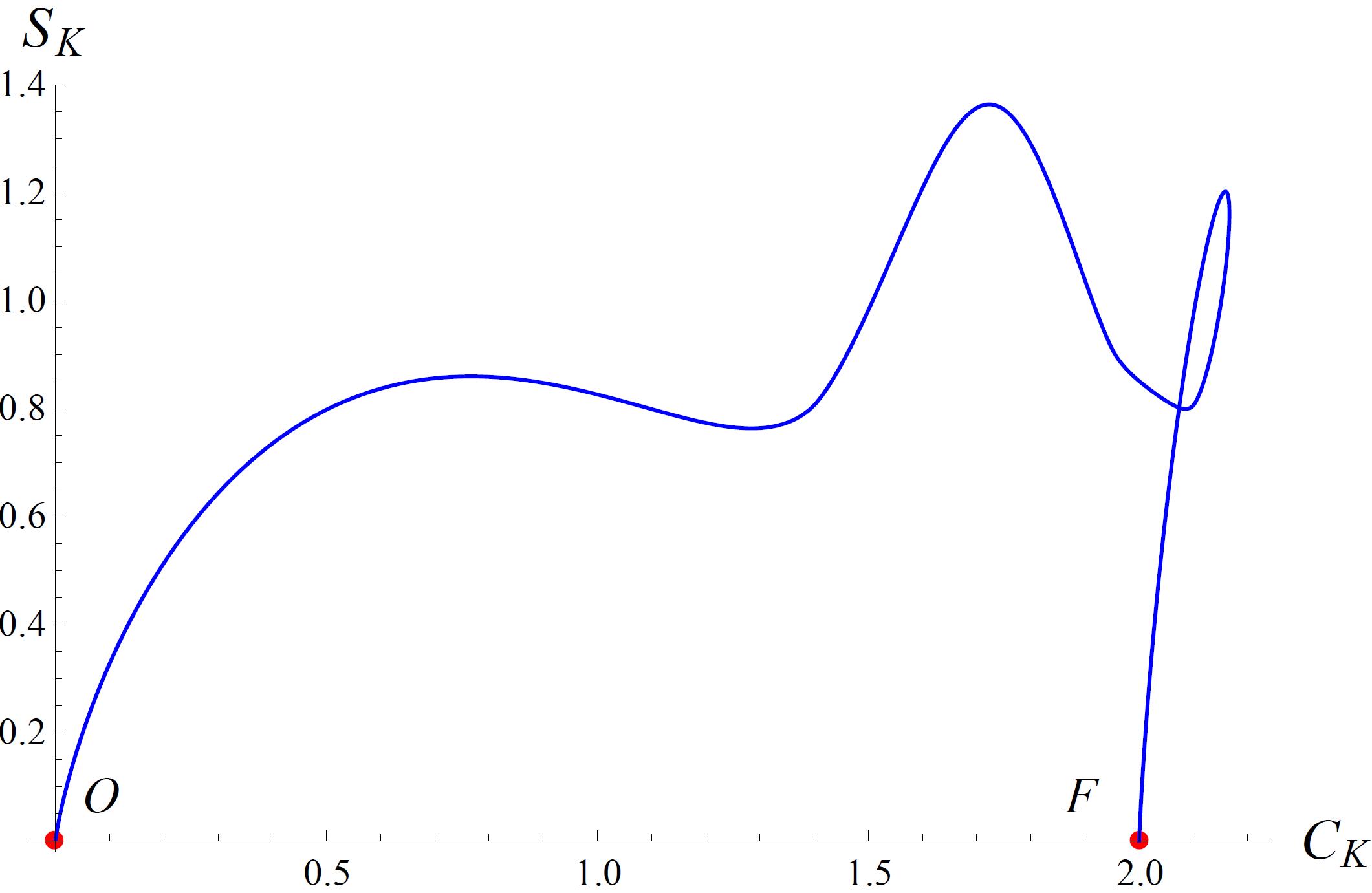}
  \includegraphics[width=140pt]{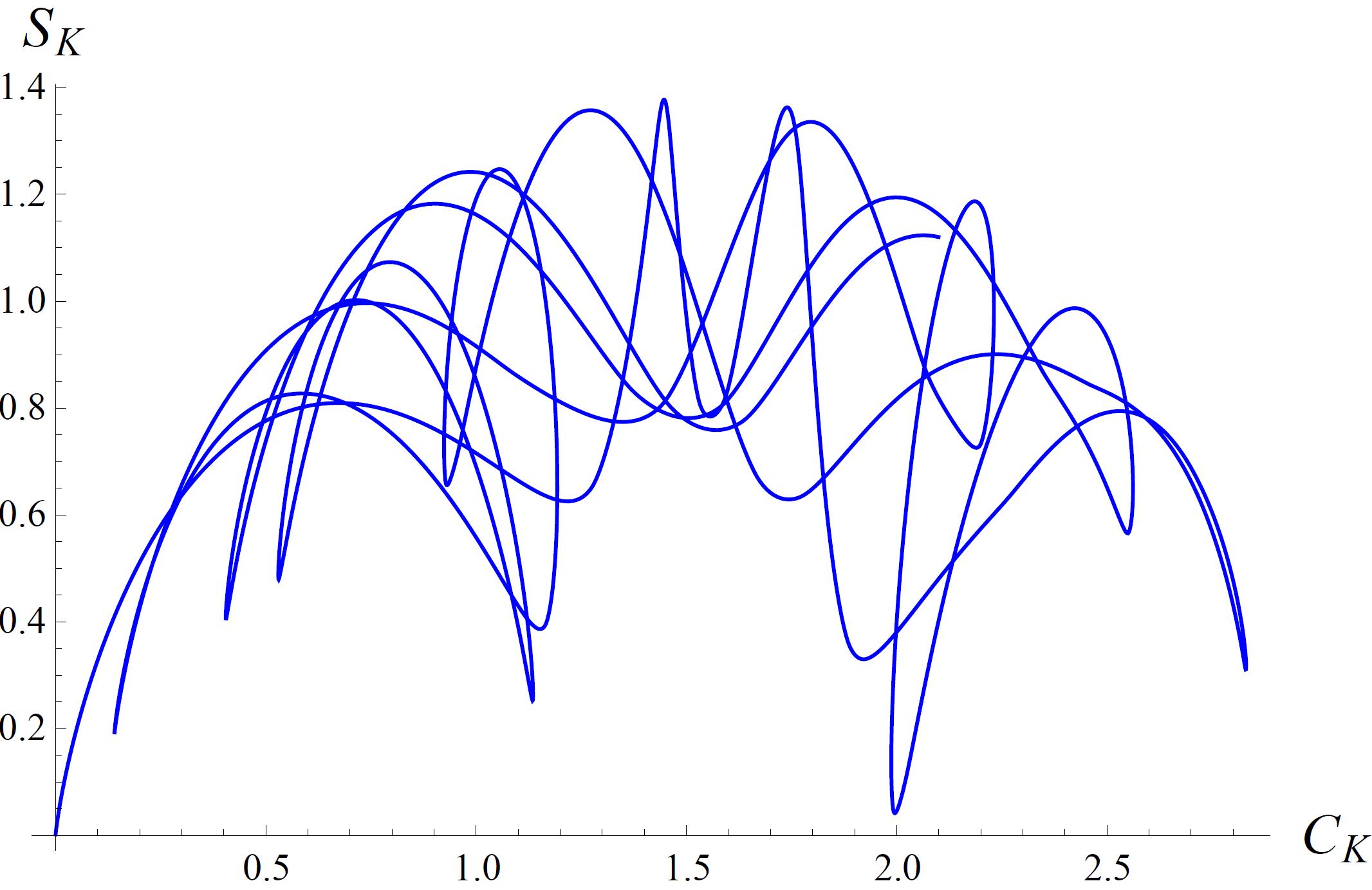}
  \caption{Trajectories on $C_K$-$S_K$ plane for finite chains. Left panel: $K=1$. Middle panel: $K=3$ and $\omega_2=2\omega_1$. In both cases, the particle moves periodically between $O$ and $F$. In the first half period of $C_K$, $t\in [0\,,T_C/2]$, it moves to the right, from $O$ to $F$ whereas in the next half period $t\in [T_c/2\,,T_C]$, it moves in the opposite direction, similar to a harmonic oscillator. Right panel: $K=3$ and $\omega_2=\sqrt{3}\omega_1$. The motion of the particle is not periodic and the trajectory becomes more and more complex as time increases. Here we have chosen $t\in [0\,,20]$. }
  \label{finite}
\end{figure}

Let us consider several examples. The first is $K=1$: $b_1=\omega$ and $b_n=0$ otherwise. The auto-correlation function is simply a cosine function $\varphi_0(t)=\cos{(\omega t)}$ and $\varphi_1(t)=\sin{(\omega t)}$. One easily finds
\bea\label{finite1}
&& C_K(t)=\sin^2(\omega t)\,,\nn\\
&& S_K(t)=-\cos^2(\omega t)\,\mm{ln}{\cos^2(\omega t)}-\sin^2(\omega t)\,\mm{ln}{\sin^2(\omega t)}\,.
\eea
It is clear that the both are periodic in times. The minimal period is $T=\pi/\omega$ for $C_K$ and $T=\pi/2\omega$ for $S_K$. The functional relation $S_K(C_K)$ is shown in the left panel of Fig. \ref{finite}. We may think of the time evolution as a particle moving on the $C_K$-$S_K$ plane. In the first half period of $C_K$, $t\in [0\,,T_C/2]$, the particle moves to the right, starting at $O$ and stopping at $F$, giving rise to a finite path. In the next half period $t\in [T_c/2\,,T_C]$, the particle still moves along the same trajectory but in the opposite direction exactly: it goes from $F$ to $O$. This is very similar to a harmonic oscillator. The same behavior will be repeated as time increases.

The second example is $K=3$:
\be b_1^2=\fft{\omega_1^2+\omega_2^2}{2}\,,\quad b_2^2=\fft{(\omega_1^2-\omega_2^2)^2}{2(\omega_1^2+\omega_2^2)}\,,\quad b_3^2=\fft{2\omega_1^2\omega_2^2}{\omega_1^2+\omega_2^2}      \,.\ee
The auto-correlation function is given by $\varphi_0(t)=\big(\cos(\omega_1 t)+\cos(\omega_2 t) \big)/2$. If the ratio $\omega_2/\omega_1$ is rational, $\varphi_0(t)$ will be periodic as well as the K-complexity and K-entropy. The situation is quite similar to the previous case except some slight differences, as shown in the middle panel of Fig. \ref{finite}. However, when $\omega_2/\omega_1$ is irrational, the auto-correlation function will not be periodic. The particle does not move on a fixed path and the trajectory on the plane will become more and more complex in the time evolution, see the right panel of Fig. \ref{finite}. These are general features for odd $K$.

For even $K$, as previously emphasized, the results can be obtained from $(K+1)$-case by setting one of the $L$ frequencies $\omega_\ell$ equal to zero. For example, for $K=2$ case, the auto-correlation function is given by $\varphi_0(t)=a_0+a_1\cos{(\omega t)}$, where $a_0+a_1=1$. Of course, the time evolution is periodic except that $W=\infty$ because of the constant term $a_0\neq 0$. For $K=4$ case, $\varphi_0(t)=a_0+a_1\cos{(\omega_1 t)}+a_2\cos{(\omega_2 t)}$, where $a_0+a_1+a_2=1$. Whether the time evolution is periodic or not depends on the ratio $\omega_2/\omega_1$ is rational or irrational. In any case, the trajectory on the $C_K$-$S_K$ plane shows the same features as the odd $K$ case qualitatively. Therefore, we safely conclude that for reversible process characterized by $W=0$ or $W=\infty$, the functional relation $S_K\sim \log{C_K}$ at long times does not hold any longer.


\section{Conclusion and discussion}


In this paper, we study quantum information quantities: K-complexity $C_K(t)$ and K-entropy $S_K(t)$ in operator growth. We are trying to search their interesting features which may diagnose irreversibility of the process. We have studied a variety of systems with emergence of dissipative behaviors, including chaotic ones and integrable theories.
Our main result is for irreversible process, the two quantities enjoy a logarithmic relation (\ref{skchaotic}) to leading order at sufficiently long times, where the proportional constant is up bounded as $\tilde\eta\leq 1$. It should be emphasized that in practical calculations we choose a certain inner product in the operator space and take the thermodynamic limit. However, the logarithmic relation only depends on the asymptotic behavior of Lanczos coefficients $b_n$. Hence choosing different inner product or considering finite temperatures will not change the relation if the asymptotic behavior of $b_n$ takes the same asymptotic form as those studied in the paper. Inspired by similarity of the relation to the Boltzmann formula in statistical mechanics, we propose that the relation is a sufficient condition for irreversibility of operator growth, irrespective of the choice of dynamics. This is true as far as we can check although we cannot prove it analytically. Physical consequences of the relation deserves further investigations.
\begin{table}[h]
\centering
\begin{tabular}{|c|c|c|c|c|c|c|c|}
  \hline
  model  &  chaotic     & $1d$-chaotic        &     integrable                 &     unknown     &  unknown    &   bounded     \\\hline
  $b_n$ & $\alpha n $   & $\alpha n/\mm{ln}{n} $  &     $ \alpha n^\delta $       &   $\alpha n^\delta (\mm{ln}{n})^\pm$      &  $\alpha\,\mm{ln}{n} $  &  $b$   \\ \hline
  $C_K$ & $ e^{2\alpha t} $  & $ e^{\sqrt{4\alpha t}} $ & $ (2\alpha t )^{\gamma} $  & $ (2\alpha t )^{\gamma} \,\mm{ln}^{\pm\gamma}{(2\alpha t)} $   &$2\alpha t\,\mm{ln}{(2\alpha t)}$ & $2bt$ \\ \hline
  $S_K$ & $ 2\alpha t $  & $ \sqrt{4\alpha t} $ & $ \mm{ln}{(2\alpha t )} $  & $ \mm{ln}{(2\alpha t )}$   & $ \mm{ln}{(2\alpha t )}$ &  $ \mm{ln}{(2b t )}$ \\ \hline
\end{tabular}
\caption{Asymptotic growth of Lanczos coefficient $b_n$ and the leading long time dependence of K-complexity and K-entropy. The constants $\alpha$ and $b$ have dimension of energy and $\gamma=\fft{1}{1-\delta}$, $0<\delta<1$. The last column corresponds to the case with bounded support, of which a particularly interesting example is chaotic systems in the post-scrambling regimes. The fifth/sixth columns can be viewed as integrable models/bounded case with logarithmic corrections. }
\label{ckbn}
\end{table}

In addition to the logarithmic relation, behavior of K-complexity itself is also closely related to irreversibility. In table 1, we summarize asymptotic growth of Lanczos coefficient $b_n$
and the leading long time dependence of $C_K$ (and $S_K$). It is intriguing to observe that the former is somehow positively related to the latter: when $b_n$ grows faster asymptotically, $C_K$ grows faster in the time evolution (in a one-to-one mapping) as well. If this is correct, from the long time behavior of $C_K$, one can read off the asymptotic behavior of the Lanczos coefficients. The underlying relation between $C_K$ and emergence of dissipative behavior certainly deserves further investigations.

\section*{Acknowledgments}

Z.Y. Fan was supported in part by the National Natural Science Foundations of China with Grant No. 11805041 and No. 11873025.

\end{document}